\documentclass[graybox]{svmult}

\usepackage{type1cm}        % activate if the above 3 fonts are
\usepackage{makeidx}        % allows index generation
\usepackage{graphicx}       % standard LaTeX graphics tool
\usepackage{multicol}       % used for the two-column index
\usepackage[bottom]{footmisc}% places footnotes at page bottom

\usepackage{newtxtext}       % 
\usepackage{newtxmath}       % selects Times Roman as basic font

\usepackage{tabularx}
\usepackage{multicol}
\usepackage{multirow}
\usepackage{booktabs}
\usepackage{graphicx}
\usepackage{enumitem}
\usepackage{url}

\newcommand{\multicell}[2][t]{\begin{tabular}[#1]{@{}l@{}}#2\end{tabular}} % allow linebreaks in cell

\usepackage{tikz}
\newcommand\copyrighttext{%
  \footnotesize This is a preprint of the following chapter: Franziska Dobrigkeit, Christoph Matthies, Ralf Teusner and Michael Perscheid, \emph{Joining Forces: Applying Design Thinking Techniques in Scrum Meetings}, published in Design Thinking Research, edited by Christoph Meinel, Larry Leifer, 2021, Springer reproduced with permission of Springer Nature Switzerland AG. The final authenticated version is available online at: \url{https://doi.org/10.1007/978-3-030-76324-4_17}.
  }
\newcommand\copyrightnotice{%
\begin{tikzpicture}[remember picture,overlay]
\node[anchor=south,yshift=10pt] at (current page.south) {\fbox{\parbox{\dimexpr\textwidth-\fboxsep-\fboxrule\relax}{\copyrighttext}}};
\end{tikzpicture}%
}

\makeindex             % used for the subject index
\begin{document}

\title*{Joining Forces: Applying Design Thinking Techniques in Scrum Meetings}
\author{Franziska Dobrigkeit, Christoph Matthies, Ralf Teusner, Michael Perscheid}
\institute{Franziska Dobrigkeit \at Hasso Plattner Institute \email{franziska.dobrigkeit@hpi.de}
\and Christoph Matthies \at Hasso Plattner Institute \email{christoph.matthies@hpi.de}
\and Ralf Teusner \at Hasso Plattner Institute \email{ralf.teusner@hpi.de}
\and Dr. Michael Perscheid \at Hasso Plattner Institute \email{michael.perscheid@hpi.de}}
\maketitle

\copyrightnotice

\abstract{The most prominent Agile framework Scrum, is often criticized for its amount of meetings.
These regular events are essential to the empirical inspect-and-adapt cycle proposed by Agile methods.
Scrum meetings face several challenges, such as being perceived as boring, repetitive, or irrelevant, leading to decreased cooperation in teams and less successful projects.
In an attempt to address these challenges, Agile practitioners have adopted teamwork, innovation, and design techniques geared towards improving collaboration.
Additionally, they have developed their own activities to be used in Scrum meetings, most notably for conducting retrospective and planning events.

Design thinking incorporates non-designers and designers in design and conceptualization activities, including user research, ideation, or testing.
Accordingly, the design thinking approach provides a process with different phases and accompanying techniques for each step.
While these techniques are often not new, they are revised and customized for teams with little design experience.
These design thinking techniques can support shared understanding in teams and can improve collaboration, creativity, and product understanding.
For these reasons, design thinking techniques represent a worthwhile addition to the Scrum meeting toolkit and can support Agile meetings in preventing or countering common meeting challenges and 
achieving meeting goals.

This chapter explores how techniques from the design thinking toolkit can support Scrum meetings from a theoretical and practical viewpoint.
We analyze Scrum meetings' requirements, goals, and challenges and link them to groups of techniques from the design thinking toolkit.
In addition, we review interview and observational data from two previous studies with software development practitioners and derive concrete examples.
As a result, we present initial guidelines on integrating design thinking techniques into Scrum meetings to make them more engaging, collaborative, and interactive.}

\section{Introduction}\label{sec:introduction}
Agile software development processes, especially Scrum and Scrum-hybrids (i.e. Scrumban or Scrum/XP), are standard practices in modern software companies.
Team meetings employed in these methods facilitate empirical process control, create regularity, and decrease the demand for additional unplanned or unstructured sessions~\cite{Schwaber2017,rubin12}.
The literature regarding Scrum provides detailed descriptions of each meeting and \textit{what} should be accomplished in them~\cite{Schwaber2017}.
However, few prescriptions or best practices on \textit{how} these goals should be achieved exist.
Consequently, teams face several challenges concerning the facilitation and perceived effectiveness of meetings including:
\begin{itemize}
    \item Not all meetings are equally relevant for all participating team members, leading to a decrease in participant engagement~\cite{Akif2012}.
    \item Planning and review meetings may be perceived as a waste of development time and as too ``simple''~\cite{Cho2008}.
    \item Regular meetings featuring identical structures create monotony which can negatively affect outcome quality~\cite{kua2013}.
    \item A lack of follow-through regarding meeting outcomes leads to a feeling that ``the same things are discussed over and over''~\cite{Przybyek2017}.
    \item The transparency afforded by Agile approaches can be abused leading to feelings of being exposed or inadequate in team members~\cite{Conboy2011}.
\end{itemize}

\noindent
Knowledge of these common meeting issues allows addressing, resolving, and preventing problems before they become hindrances to team collaboration~\cite{Cho2008}.
Researchers and practitioners have proposed and collected various activities and best practices for different phases and contexts of the Scrum process.
Much of these focus on Retrospective~\cite{Baldauf2018,jovanovic2016,Ng2020} or Sprint Planning meetings, e.g., the common \emph{Planning Poker} in
Scrum~\cite{rubin12,Haugen2006}, but other activities exist~\cite{hohmann2006,Gray2010}. 

Design thinking (DT) has emerged as an approach that can support software development by providing interactive and engaging techniques for collaborative problem analysis and solution development.
It can improve creativity, product understanding, collaboration, and empathy towards customers and the team~\cite{plattner_DT_2020}.
Employing DT as a preceding phase to Agile software development is a popular scenario~\cite{Rafael2019}.
However, in line with previous studies~\cite{Pereira2018}, our research has also shown that the application of DT in later stages of the development cycle is valuable for development teams~\cite{dobrigkeit_design_2019}. 
Developers especially value the change in working style and the required customer-centered mindset.
As such, DT techniques provide an opportunity to add further useful activities to the Scrum meeting toolbox. They can help address common challenges, achieve meeting requirements for attendees, or aid in reaching meeting outcomes that have the support of all team members. We, therefore, explore how techniques from DT can be applied to support Scrum meetings.

\section{Background and Related Work}
\label{sec:background}

In this Section, we briefly describe the key concepts of Scrum and DT as well as existing research into their activities and techniques.
Additionally, we describe existing approaches to combine Agile Software Development and DT.

\subsection{Scrum}
Scrum is an Agile process framework for managing work on complex products in short iterations (\emph{Sprints}), with an emphasis on software development~\cite{Schwaber2017}.
It defines the different roles of a Scrum team, the rules to follow, the artifacts to create and the meetings that steer and organize a Sprint~\cite{Schwaber2017,rubin12}. 
A Scrum team consists of the Product Owner (PO), the Scrum Master (SM), and the Development Team.
The PO is responsible for managing the product vision and the \emph{Product Backlog}, which contains the required work items needed to achieve the vision.
The SM facilitates the Scrum meetings and supports the team in overcoming issues.
The development team develops the (software) product.
Work happens in an iterative, incremental manner, carried out in cycles called Sprints~\cite{ScrumPrimer2012}.
There are five main Scrum meetings that facilitate team cooperation and collaboration~\cite{rubin12}:
\begin{itemize}
    \item \textit{Backlog Refinement}, where work items are prepared
    \item \textit{Planning}, where the work items for the next Sprint are decided
    \item \textit{Daily Scrum}, provides a regular team status update
    \item \textit{Review}, concerned with inspecting the product and collecting feedback
    \item \textit{Retrospective}, focuses on process improvement
\end{itemize}

We provide a detailed description of each meeting with its requirements and outcomes in Section \ref{sec:scrum_meetings}.

\subsubsection{Meeting Challenges}
Studies on the subject of the challenges involved in running Scrum meetings have identified multiple issues of relevance.
The literature notes that Scrum events are not equally relevant for all participating team members, leading to a decrease in engagement~\cite{Akif2012}.
Similarly, meetings such as Sprint Planning and Review, which were perceived by participants as ``simple'', have been described as a waste of valuable development time~\cite{Cho2008}.
If Scrum meetings are repeatedly held in the same fashion and with similar structures, apathy and monotony felt by attendees can negatively affect the quality of meeting outcomes~\cite{kua2013}.
In case studies of multiple teams, Retrospective meetings were judged to be ineffective by participants, as the same issues were repeatedly discussed in subsequent meetings~\cite{Przybyek2017}.
Furthermore, the downsides and drawbacks of the full transparency and visibility provided by Scrum events have been highlighted, such as exposing developer shortcomings publicly.
This can prove to be counterproductive, as team members can be made to feel inadequate or can lead to unhealthy environments where this transparency is abused~\cite{Conboy2011}.

The prescriptive or ``dogmatic'' nature of Scrum events has been criticized, e.g., regarding the Daily Scrum meeting, which is strictly time-boxed and should ideally always be performed every day at the same time with the entire team.
These requirements can break down in real-world circumstances featuring distributed teams, meeting setup times, flexible working schedules, and varying team sizes~\cite{Meyer14}.

\subsubsection{Existing Activities for Scrum Meetings}
Various activities and best practices have been proposed in the related literature for different phases and contexts of the Scrum process to address common Scrum meeting challenges.
In particular, Scrum process facilitators have introduced tools and meeting agendas that help in achieving the main goals of Scrum meetings, e.g., for the Retrospective (focusing on creative prompts for feedback collection) and Sprint Planning (focusing on collecting effort estimations),

Much previous research is available on Retrospectives, and the activities that teams can employ during these meetings and studies continue to be published~\cite{Ng2020}.
Retrospective meetings have been in use by teams since before Agile methods became widespread.
Team activities or ``games'' that participants play to keep sessions from becoming stale and repetitive have been used in Retrospectives since their inception~\cite{kerth2000}.
These interactive games are designed to encourage the required reflection and team collaboration needed for the meeting.
Derby and Larsen describe the purpose of Retrospective activities as helping a team ``think together''~\cite{Derby2006}.
Retrospective activities have been shown to positively impact the creativity, involvement, and communication of team members~\cite{Przybyek2017}.

Sprint Planning and Backlog Refinement meetings are also closely associated with best practices in agile processes.
These activities are concerned with the requirement of producing effort estimates for work items.
Unstructured group estimation processes pose the risk of being highly influenced by company politics, group pressure, or dominant personalities, which may reduce estimation performance~\cite{Haugen2006}.
To allow teams to come up with shared estimates while reducing bias-related anti-patterns, tools such as \emph{Planning Poker} were introduced~\cite{Ralph2013}.
Through the use of structured meeting agendas and categories of answers, these activities can help ensure that all team members participate and that their opinions are heard, regardless of group influence~\cite{Haugen2006}.

Scrum practitioners have recommended different activities for use in Scrum meetings, but few of these have garnered widespread attention and adoption.

\subsection{Design Thinking}
The term Design Thinking is used ambiguously and can describe a) a cognitive style of working dominant with designers, b) the idea of including design practices into all parts of an organization or c) an approach to creating new and innovative products and services.
In this article, we understand DT as an approach to new product and service development, as it is taught at Stanford or the HPI School of Design Thinking Potsdam.
In this approach, the design process and methods from design and other disciplines have been remolded in a way that non-designers can apply them to successfully create new products and services.
Within this understanding of DT, the DT process is described as an iterative and flexible multi-step process with recommended techniques for each step.
While authors do not agree on the number of steps, the steps of different processes can be mapped to a) steps that focus on learning about the problem and the user, b) ideating on a solution, and c) prototyping and testing this solution with users.
Detailed descriptions of these processes are available~\cite{brown_design_2008, thoring_understanding_2011, wolbling_design_2012, brenner_design_2016}.
The recommended techniques accompanying each process step, provide practitioners with multiple ways to facilitate each step (cmp.~\cite{ideo_ideo_2003, hanington2012universal, kumar_101_2012}).
We give a detailed overview of such techniques in Section~\ref{sec:dt_techniques}.

Research on DT found that its advantages include increased collaboration, empathy, product differentiation, and cost savings due to reduced redesign work and shorter lead time to development~\cite{carlgren_design_2014}.

The working style for most DT techniques is very interactive and research with development teams suggests that developers welcome this variety in work style when applying DT. Additionally, research suggests that developers value and implement DT techniques when they have the required knowledge \cite{dobrigkeit_design_2019} and that developers can increase empathy towards users and team members, as well as shared product understanding and collaboration \cite{plattner_DT_2020}.

\subsection{Design Thinking and Agile Software Development}
Research on how to integrate DT and software development suggests different approaches.
Lindberg et al. present four models of how DT can support software development \cite{lindberg_perception_2012}.
We previously suggested adding two more models and expanding the list to six~\cite{dobrigkeit_design_2019}.
Table~\ref{tab:DT-ITmodels} provides an overview of these six integration possibilities.

\begin{table}[htbp]
\center
\caption{Overview of models integrating Design Thinking and software development processes (adapted from \cite{dobrigkeit_design_2019}).}
\label{tab:DT-ITmodels}
\begin{tabularx}{\columnwidth}{lX}
    \toprule
    \textbf{Name} & \textbf{Description}\\
    \midrule
    Split project model & DT as separate project before development\\
    Overlapping teams model & DT as initial process phase with one or more development team members participating \\
    Unified project model & DT as initial process phase and a large overlap of DT team and development team \\
    Continuous DT model & Smaller DT phases are regularly integrated into agile development cycles\\
    Ad-hoc DT model & DT as supporting tool that development teams can apply in form of DT workshops or DT phases if required\\
    Toolbox model & Methods developers can use to overcome problems they can't solve with IT methods\\
    \bottomrule
\end{tabularx}
\end{table}

In line with these proposals, researchers have suggested different integration approaches covering the range of different models.
Running DT as a preceding phase to agile development, as suggested in the first three models seems the most common form: the Integrated Design Thinking and Lean Development Approach \cite{hildenbrand_intertwining_2012}, the Nordstrom Innovation Lab Process \cite{grossman-kahn_skip_2012}, InnoDev \cite{plattner_innodev_2018} or Converge \cite{marcus_software_2015}.

Continuous approaches of applying DT and software development are proposed by the Integrated Design Thinking and Agile Framework for Digital Transformation \cite{marcus_integrated_2016}, the Human-centered Agile Workflow \cite{ahram_introducing_2018} and DT@XP \cite{sohaib_integrating_2018}, which suggest regular DT workshops or phases as part of each development cycle.
Ad-hoc DT implementations in later stages of the development cycle are proposed by InnoDev \cite{plattner_innodev_2018} and Converge \cite{marcus_software_2015}.
These approaches propose spontaneous DT workshops with low planning overhead, when new features are added to the product, when tackling identified blockers, or supporting agile meetings.
Toolbox approaches that implement these concepts were previously suggested~\cite{plattner_DT_2020, pedersen_ux_2016}. 

Adding DT techniques to the toolbox of Scrum meeting facilitators would be in line with approaches that follow the Ad hoc DT Model or the Toolbox Model.

\section{Methodology}
We performed a two-step study to explore how techniques from the DT toolkit can be applied to address the challenges of Scrum meetings.
First, we compared the requirements, challenges, and existing method recommendations from the Scrum literature with various methods described in the DT literature. 
In a following step, we supported this theoretical comparison with empirical evidence and concrete examples collected from a meta-analysis.
In this meta-analysis, we reviewed data collected in two former studies in which the topic of \emph{DT techniques in Scrum meetings} had emerged.

The first original study \emph{(OS-1)} was a ten-month longitudinal study in a global software company with a team of three Scrum teams who were experienced in both Scrum and DT.
In this study, we explored when, where, and how DT was used during agile development.
We observed several of the team meetings during the ten month period and interviewed eleven team members regarding their experiences using DT in their team.
Additionally, we interviewed five employees from two further teams in the same company.
This study and its results are further described in \cite{dobrigkeit_design_2019}.

In the second original study \emph{(OS-2)}, we facilitated workshops introducing DT tools in agile development teams from six different companies.
In this study, we explored the comprehensibility and applicability of a DT toolbox to everyday agile development with DT beginners.
To this end, we prepared a collection of worksheets for twelve different DT tools, which we validated with six different companies.
We observed one company in detail over a period of twelve weeks, in which they were introduced to one method each week and were asked to apply that method at least twice during the week.
At the end of each week we interviewed the team in a group interview and asked all team members to fill out a method survey. 

The five other companies were introduced to our toolkit in a workshop format in which we introduced them to three of the methods from our toolkit and afterwards conducted a group interview. The interview focused on finding out the experiences made within the workshop and if, where, and when team members would make use of these methods again.
This study and its results are described in detail in a previous publication~\cite{plattner_DT_2020}.

\subsection{Research Context}

As the focus of OS-1 was on understanding where, when, and how DT supports agile development, we chose the global software company because it had embraced DT as well as agile development practices for several years. The three teams participating in OS-1 were chosen because they were reported to be using DT in their daily work and accordingly all interviewees in this study had prior experience with DT, albeit to a different extent.

The six companies chosen for OS-2 had little or no prior experience with DT, with the focus of this study being to explore the applicability of a toolbox for DT novices.

The participants of these two studies featured a broad range of DT experience.
Table~\ref{table:teams} provides an overview of the teams that we observed and interviewed for this study.

\begin{table}[!t]
\caption{Teams interviewed and observed for this meta analysis}
\label{table:teams}
\renewcommand{\arraystretch}{1.3}
\centering
\begin{tabularx}{\columnwidth}{llcllX}
\toprule
\textbf{\multicell{Original\\Study}} & \textbf{\multicell{\\Team}} & \textbf{\multicell{\\\# People}} & \textbf{\multicell{DT\\Exp.}} & \textbf{\multicell{Scrum\\Exp.}} & \textbf{\multicell{Short\\description}} \\
\midrule
OS-1 & Team A & 25 & \multicell{medium\\to high} & high & German-based team, incl designers developers and managers of a national software company developing a software for real estate management.\\
OS-1 & Team B & 3 & high & high & US-based team, 1 designer, 1 developer and 1 manager of a national software company that develops proof-of-concept prototypes for new technologies or unsolved customer problems.\\
OS-1 & Team C & 2 & high & high & US-based team, 1 designer and 1 manager of a national software company that acts as DT consultants to company customers, supporting them during their own DT efforts through training, coaching, and co-creation.\\
OS-2 & Team D & 10 & \multicell{low to\\medium} & high & Team of programmers and designers from a medium sized, German software development company.\\
OS-2 & Team E & 6 & \multicell{low to\\medium} & high & Team of programmers working in data analysis, artificial intelligence, machine learning and computer vision from a German research organization.\\
OS-2 & Team F & 8 & \multicell{low to\\medium} & high & Team incl. a sales-manager, a designer and a mixed group of front-end and back-end developers from a German company developing team collaboration and security solutions.\\
OS-2 & Team G & 9 & \multicell{low to\\medium} & high & Software developers of different teams from a German company developing project management software.\\
OS-2 & Team H & 8 & \multicell{low to\\medium} & high & Diverse team incl. UX designers, Agile coaches and developers of a German company developing business applications.\\
\bottomrule
\end{tabularx}
\end{table}

\subsection{Data Collection and Analysis}
For the theoretical comparison, we extracted the requirements and outcomes of each of the Scrum meetings as well as existing activities recommended for these meetings from the seminal literature.
We concentrated on the descriptions by the Scrum originators~\cite{Schwaber2017, Sutherland2007} and highly-referenced introductory literature that presents and describes the basic Scrum events~\cite{rubin12,ScrumPrimer2012,Kniberg2015,Meyer14}.
We compared these requirements with techniques frequently mentioned in the DT literature to create a mapping of Scrum meetings and DT techniques that can support them.

For the meta-analysis we re-analyzed the data collected during OS-1 and OS-2. 
In OS-1, we wrote observation notes during various Scrum meetings of the observed Team A over a period of ten months and recorded and transcribed eleven interviews with members of that team as well as five additional interviews with members of two further teams (Team B and C) from the same company. 
In OS-2, we collected observation notes over a period of twelve weeks within Team D and during five workshops with teams E-H. Additionally, we recorded and transcribed twelve group interviews in Team D (one after each week) and one group interview with each of the other four teams following their workshops.
In this meta-analysis we iteratively coded the available materials with a focus on the use of DT during Scrum meetings. We thus derived concrete examples to support our initial mapping. As a result we created initial guidelines on how to integrate DT techniques in each of the Scrum meetings for Scrum Masters and teams that want to improve their meetings' level of interactivity and collaboration or simply desire more variety.
\section{Goals and Requirements of Scrum Meetings}
\label{sec:scrum_meetings}
The Scrum meetings, sometimes referred to as \emph{ceremonies}~\cite{Meyer14}, conducted by teams during a development iteration form the core of the Scrum software development method.
These meetings mark points in the process where team members regularly discuss iteration progress and process aspects as well as work on development artifacts.
The Scrum Guide describes the purpose of these meetings as creating ``regularity and to minimize the need for meetings not defined in Scrum''~\cite{Schwaber2017}.
Schwaber and Sutherland point out that all meetings are essential and that not including a prescribed Scrum meeting results in ``reduced transparency'' and a ``lost opportunity to inspect and adapt''~\cite{Schwaber2017}.
In the following, we present and characterize the five central Scrum meetings, with a particular focus on their requirements and outcomes.

\subsection{Product Backlog Refinement}
In the \emph{Product Backlog Refinement} meeting, also referred to as \emph{Backlog Grooming}~\cite{Kniberg2015}, the team focuses on preparing product backlog work items for the following one or two Sprints~\cite{ScrumPrimer2012}.
Refinement includes activities such as analysis of requirements, adding required details to work items, estimating implementation effort, or splitting large work items~\cite{Schwaber2017}.
Furthermore, outdated or redundant items should be removed from the product backlog to keep its components as clear and actionable as possible.
While the refinement of work items can form part of a Planning meeting, depending on teams' preferences and context, most of this work should be done before the Planning meeting~\cite{Kniberg2015}.
Separating these meetings allow the subsequent Planning to take advantage of well-analyzed and estimated items, making it more efficient~\cite{ScrumPrimer2012}.

Many of the details of how Product Backlog Refinement is conducted are up to the team.
Scrum does not feature a defined time box or even a frequency for the meeting.
However, the Scrum Guide notes that the meeting should not consume ``more than 10\% of the capacity of the Development Team''~\cite{Schwaber2017}.

\smallskip
\noindent\textbf{Attendees:} Development Team, Scrum Master, Product Owner.

\noindent\textbf{Point in Process:} Custom, depending on the Scrum implementation.

\noindent\textbf{Requirements:} A shared understanding of both the technical effort to implement a solution and the business value it delivers. This allows establishing cost estimates and enables discussions on priorities.

\noindent\textbf{Outcome:} A set of well-understood, clearly described, thoroughly analyzed, and estimated set of user stories in the Product Backlog~\cite{ScrumPrimer2012}.

\subsection{Sprint Planning}
A \emph{Sprint Planning} meeting marks the start of a new development iteration in Scrum.
They are used to prepare for the upcoming Sprint and are typically divided into two parts: the first concerned with \emph{what} to build, the second with the specifics of \emph{how} to implement it~\cite{ScrumPrimer2012}.
The Product Owner communicates their vision of what is most critical regarding the project's next steps.
Together with the Development Team, the PO discusses the highest-priority items (from the Product Backlog), which represent those features that are most interesting for the near future.
The contexts and details for these high-priority items are reviewed by team members, providing the rationale and a shared understanding of the value inherent in the planned features.

Based on these discussions and insights, the team may then define a shared, relatively stable \emph{Sprint Goal}~\cite{ScrumPrimer2012} and the next items to be worked on~\cite{rubin12,Sutherland2007}.
The Sprint Goal summarizes the objective that the team aims at achieving within the Sprint, which ideally has a cohesive theme.
It is based on the latest product increment, the projected capacity, and the past performance of the team~\cite{ScrumPrimer2012}.
To forecast the functionality that can be developed during the Sprint, the entire Scrum team collaborates to understand the work needed to achieve the Sprint Goal~\cite{Schwaber2017}.
The Planning meeting provides an opportunity to flesh out further details of the work items planned for the first days of the Sprint, e.g., by creating additional subtasks for work items or bugs.
When effort estimates are not yet available (e.g., as work items were recently updated), the team works together to provide these assessments.
Considering the priority and effort estimates, the Development Team decides on the amount as well as the specific tasks it will complete in the upcoming Sprint.
The Development Team has the ability and responsibility to additionally include relevant work items with potentially lower priority, e.g., in the case of dependencies~\cite{ScrumPrimer2012}.

By the end of the Sprint Planning meeting, the Development Team should have formed a consensus on the plan of action and should be able to explain how it intends to accomplish the Sprint Goal and build the anticipated product increment~\cite{Schwaber2017}.

\smallskip
\noindent\textbf{Attendees}: Development Team, Scrum Master, Product Owner (optional for breaking down items into tasks).

\noindent\textbf{Point in Process:} At the beginning of a Sprint.

\noindent\textbf{Requirements:}
The Product Owner representing the customer and the development team must share a mental model and a deep (technical) understanding of the work ahead and the business value it presents~\cite{Conboy2011}.

\noindent\textbf{Outcomes:}
\begin{itemize}[noitemsep,topsep=0pt]
    \item A coherent Sprint Goal, summarizing the development work to be done.
    \item A list of items to be worked on during the next Sprint, ideally including the highest priority ones.
    \item A plan of how to deliver the planned functionality in a product increment.
\end{itemize}

\subsection{Daily Scrum}
The \emph{Daily Scrum} meeting is a short team meeting to enable collaboration, in which the Development Team inspects and synchronizes their work as well as outline the next aims on a daily basis~\cite{Schwaber2017}.
The primary justification for a daily meeting during a Sprint is the general Agile principle that direct contact is critical to project success~\cite{Meyer14}.
In the Daily Scrum, progress toward the specified Sprint Goal and the remaining work items of the Sprint Backlog are reviewed, with a particular focus on identifying impediments to the team's productivity~\cite{ScrumPrimer2012}.
Necessary In-depth discussions of identified issues are postponed~\cite{rubin12} and the Scrum Master helps team members resolve obstacles in the following steps~\cite{ScrumPrimer2012,Sutherland2007}.
Daily Scrum meetings are designed to be time-boxed to 15 minutes, to improve communication, and to possibly eliminate further meetings by facilitating quick decision-making~\cite{Schwaber2017}.

The structure of the Daily Scrum meeting is decided by the Development Team with a focus on producing an actionable plan for the day.
It represents an opportunity for the members of the Development Team to make realistic commitments to each other~\cite{Meyer14}.
The Scrum Guide proposes the following three questions for every participant to structure the meeting~\cite{Schwaber2017}:
\begin{itemize}[noitemsep,topsep=0pt]
    \item What did I complete yesterday that helped meet the Sprint Goal?
    \item What will I work on today to help meet the Sprint Goal?
    \item Do I see any impediments or blockers towards the Sprint Goal?
\end{itemize}

\smallskip
The Daily Scrum meeting is not intended to be used as a detailed status reporting meeting, instead it focuses on quick information sharing on what is happening across the team.
Its tone should ideally be pleasant and enjoyable while staying informative.
The meeting is also known as a \emph{Stand-up} as one of the original ideas to ensure a short session was to require participants to stand~\cite{Meyer14}.

\smallskip
\noindent\textbf{Attendees:} Development Team, (Product Owner and Scrum Master participate as developers if they are actively working on Sprint Backlog items).

\noindent\textbf{Point in Process:} Every day of a Sprint.

\noindent\textbf{Outcome:} An understanding between team members of the progress towards the Sprint Goal and the impediments that need to be tackled~\cite{Meyer14}.

\noindent\textbf{Requirements:} As the meeting is strictly time-boxed, concise recollections and summaries of recent work by attendees as well as individual reflections on encountered impediments are essential.

\subsection{Sprint Review}
Held at the end of the Sprint, the \emph{Sprint Review} meeting is used to reflect on the work of the completed Sprint~\cite{Meyer14}.
It focuses on inspecting the created product increment and eliciting feedback~\cite{Sutherland2007} and is the time to showcase the team's work.
It is, therefore, also referred to as a \emph{Sprint Demo}~\cite{Kniberg2015}, where a hands-on inspection of the real software takes place, and the Development Team gets credit for their accomplishments~\cite{Kniberg2015}.
During the meeting, the team members present the Sprint results to key stakeholders and customers and discuss progress.
The meeting provides an opportunity for the Product Owner to gain insight into the status of both the current product and the rest of the team.
Other team members gain updates from the Product Owner and the market situation~\cite{ScrumPrimer2012}.

To be demonstrable, the Sprint's work must result in a working product increment and meet the team's agreed quality bar.
In the case of software development, this might refer to a system that is integrated, tested, user documented, and ``potentially shippable''~\cite{ScrumPrimer2012}.
The Sprint Review is the second to last meeting of a development iteration.
The entire group collaborates on the next steps so that the Sprint Review provides valuable input to the Sprint Planning meetings of the next sprint~\cite{Schwaber2017}.
The Product Backlog is updated based on the completed work items or shifts in business value or the market~\cite{Schwaber2017}.

\smallskip
\noindent\textbf{Attendees:} Development Team, Scrum Master, Product Owner, other project stakeholders as appropriate.

\noindent\textbf{Point in Process:} After work on the Sprint has concluded.

\noindent\textbf{Outcome:}
A revised Product Backlog based on collected feedback~\cite{Schwaber2017}.

\noindent\textbf{Requirements:} Knowledge of the status of work items.
An understanding of how the developed functionality integrates with the rest of the product. Strategies to collect actionable feedback regarding the software increment.

\subsection{Retrospective}
The \emph{Retrospective} meeting is the last Scrum meeting of a Sprint, taking place after the Sprint Review and before the next Sprint Planning.
It is focused on inspecting and improving the executed development process of teams and is concerned with creating a plan for improvements to be enacted in the future~\cite{Schwaber2017}.
The meeting helps identify and rank aspects of the completed Sprint that have proven themselves and should be continued and those that should be discontinued~\cite{ScrumPrimer2012,Sutherland2007}.
Assumptions that led the team astray should be identified and their origins examined.
Improvement opportunities can be identified for all process aspects, including people, relationships, work process, and tools, with the overall goal of making the development process and the employed practices more effective and enjoyable~\cite{Schwaber2017}.

The Scrum method itself does not prescribe the detailed contents of Retrospectives and many activities as well as agendas for conducting the meeting have been proposed~\cite{Baldauf2018}.
By the end of the Retrospective, the team should have found a consensus on the concrete improvement actions it will attempt to implement in the next Sprint.
The most impactful improvements should be addressed as soon as possible and may be added to the next iteration's Sprint Backlog.

\smallskip
\noindent\textbf{Attendees:} Development Team, Scrum Master, Product Owner.

\noindent\textbf{Point in Process:} At the very end of a Sprint.

\noindent\textbf{Outcome:} Set of ordered action items regarding the team's work processes to be enacted in the next Sprint~\cite{Meyer14}.

\noindent\textbf{Requirements:} Detailed recollections of the enacted process of the concluded Sprint, including relevant interactions, dependencies, highlights, and encountered issues. Reflection on the severity of these items.

\section{Techniques in the Design Thinking Toolkit}
\label{sec:dt_techniques}

As described before, the DT approach uses a multitude of techniques that support each step of the DT process. These techniques originate from diverse areas, like quality management, research in creativity and design, research in communication, ethnography,
and informatics \cite{brenner_design_2016}. There is no complete list of DT techniques, and practitioners add new techniques to their toolkits from other disciplines. So while no exact definition of a DT method exists, some methods are frequently mentioned and described by researchers \cite{wolbling_design_2012, plattner_innodev_2018, carleton_playbook_2013} and practitioners \cite{ideo_ideo_2003, hanington2012universal, kumar_101_2012}. Method collections, such as \cite{ideo_ideo_2003, hanington2012universal, kumar_101_2012, carleton_playbook_2013} both from research and practice provide an overview of possible techniques and are often organized by steps in the design or DT process. For this section we propose a slightly different categorization, focusing on the tasks that can be achieved with a group of techniques. In the following,we describe the categories of techniques that are typically found in a design thinker's toolkit and give a few concrete examples for each category. Table \ref{table:methods} provides an overview of the presented categories with some examples.

\begin{table}[!t]
\caption{Categories of Design Thinking techniques and common examples}
\label{table:methods}

\renewcommand{\arraystretch}{1.3}
\centering
\begin{tabularx}{\columnwidth}{lX}
\toprule
\textbf{Method Category} & \textbf{Examples}\\
\midrule
Warm-up Techniques & Dancing in the Dark, Two Truth and One Lie, Paperclip Exercise, Marshmallow Challenge\\
\multicolumn{2}{p{0.95\columnwidth}}{\textit{Energizing group exercises designed to help participants ease into an atmosphere of teamwork.}}\\
\midrule
Unpacking Techniques & Charetting, Semantic Analysis, Mind Mapping \\
\multicolumn{2}{p{0.95\columnwidth}}{\textit{Group exercises designed to share and discuss the existing understanding and assumptions.}}\\
\midrule
Desk Research Techniques & News Search, Market Research, Technology Research \\
\multicolumn{2}{p{0.95\columnwidth}}{\textit{Research within books, published research literature or the internet,e.g. by searching for market trends, patents, news, etc.}}\\
\midrule
Field Research Techniques & Extreme User or Expert Interviews, Fly on the Wall, Shadowing, Contextual Inquiry, Immersion\\
\multicolumn{2}{p{0.95\columnwidth}}{\textit{Techniques geared towards learning directly from the user or stakeholder.}}\\
\midrule
Knowledge Sharing Techniques & Story Telling, Role Playing \\
\multicolumn{2}{p{0.95\columnwidth}}{\textit{Techniques geared towards sharing experiences, information and results from research or testing with the team.}}\\
\midrule
Knowledge Organization Techniques & Clustering, Venn-Diagram, 2x2Matrix, Timeline \\
\multicolumn{2}{p{0.95\columnwidth}}{\textit{Techniques geared towards organizing and visualizing information from research and testing.}}\\
\midrule
Knowledge Consolidation Techniques & Personas, PoV Statement, Journey Maps, Storyboards \\
\multicolumn{2}{p{0.95\columnwidth}}{\textit{Techniques geared towards condensing information from research or testing in a meaningful way usable for later process stages.}}\\
\midrule
Idea Generation Techniques & How to and How Might we Question, Brainstorming\\
\multicolumn{2}{p{0.95\columnwidth}}{\textit{Techniques that prompt the generation and sharing of ideas towards a specific prompt or question.}}\\
\midrule
Prototyping Techniques & Sketches, Role Plays, Storyboards, Wire Frames\\
\multicolumn{2}{p{0.95\columnwidth}}{\textit{Techniques that represent an aspect of an idea or a product in a meaningful way to gather feedback or allow for testing, e.g., by telling a story or presenting a schematic visual.}}\\
\midrule
Testing Techniques & Think Aloud, Concept Testing, A/B Testing, Usability Testing\\
\multicolumn{2}{p{0.95\columnwidth}}{\textit{Techniques, in which feedback or experiences of users are gathered through direct engagement with a prototype or the product.}}\\
\midrule
Feedback Techniques & I Like I wish, Feedback Capture Grid, Empathy Map\\
\multicolumn{2}{p{0.95\columnwidth}}{\textit{Techniques developed to elicit and collect structured feedback using different categories.}}\\
\midrule
Facilitation Techniques & Check-in \& Check-Out, Team Rules, Planning sessions, Reflective sessions, Voting techniques\\
\multicolumn{2}{p{0.95\columnwidth}}{\textit{Techniques usually used by coaches and facilitators to ensure effective team work.}}\\
\bottomrule
\end{tabularx}
\end{table}

\subsection{Warm-Up Techniques}
Warm-ups also called ice-breakers or energizers, are a known concept for facilitators of all kinds of team-based workshops. Warm-ups can serve a variety of functions. In the very literal sense of warming-up, these methods can help to energize participants after a longer presentation or early in the morning by providing several minutes of physical activity, e.g. Dancing in the Dark. For this activity participants block their sight, e.g. with a post it or their hats. Then the moderator starts music that is suitable for dancing and asks the participants to perform different types of dancing and movements, for example Tango, Belly Dance or a simple spin. 
In a new team or at the beginning of a workshop, warm-ups can help participants get acquainted with each other by facilitating the sharing of specific knowledge. For example, the exercise Two Truths and One Lie asks each person in the group to state three things about themselves, two of which are true and one is a lie. Afterwards, the rest of the group has to guess which one is the lie. 
During a workshop warm-up techniques can help to get the group into the right mind-set for the following exercises. If for example the next exercise for the group is to brainstorm ideas, a warm-up that practices brainstorming can help to get the creativity flowing. The paperclip exercises is such a warm-up. It asks each participant to write down as many possible usage scenarios for a big paper clip as they can over a certain time frame, e.g. 5 minutes. Similarly, if building a prototype is the next step a warm-up that requires building something gets people into the activity. For example, in the marshmallow challenge teams are competing to build the highest possible structure out of tape, spaghetti and a marshmallow. By doing so they learn to build, fail and iterate together as a team.
If at some point in time the team is in a negative mood, warm-ups can provide a fun activity that helps to create a positive atmosphere before carrying on. 
If a team is stuck, they provide distraction, making it possible to return to the task later.

\subsection{Unpacking Techniques}
Techniques used for unpacking are designed to share and discuss the existing understanding and assumptions on a topic. While doing so, teams can reach a shared understanding and uncover knowledge gaps or assumptions that have to be verified. Especially in the beginning of the process they can help to understand and analyze the design challenge. Common unpacking activities include charetting, semantic analysis or simple brainstorming techniques such as mind mapping.
Charetting is an activity that can be used for unpacking but also for ideation. During unpacking the team first brainstorms relevant users or contexts for their design challenge. In a second step the team picks one of these users and brainstorms potential issues relevant to that user. In the third step the team picks the most interesting challenge and brainstorms potential solutions. Steps 2-3 are repeated for further users. 
Following these brainstorming tasks, the team discusses what they have discovered and rephrases the design challenge to reflect the new understanding. For a semantic analysis of the given design challenge the team highlights important passages or words from the design challenge and discusses experiences, thoughts, open questions, associations and assumptions in detail. The information discussed is collected on sticky notes or the whiteboard. Thus the team explores different aspects of the challenge and reaches a common understanding. At the end of the analysis the team often rephrases the design challenge to reflect their new understanding. A less structured approach to unpacking is mind mapping, whereby the team brainstorms whatever comes to mind when thinking of the design challenge and then organizes the information in a mind map with the challenge in the center.

\subsection{Desk Research Techniques}
Desk research techniques, in contrast to field research, include analysis activities that can be done from one's desk.
These include reviewing studies, reports or patents in published research literature as well as online sources and grey literature.
Furthermore, market or technology trends can be explored usually through specific search sites on the internet.

\subsection{Field Research Techniques}
Field research techniques are designed to learn directly from users and stakeholders. Several of the techniques originate from qualitative and ethnographic research in social sciences. These techniques can be separated into interview techniques, observation techniques, and self-immersion. Interview techniques elicit information from users or stakeholders by asking them questions. Common interview techniques are extreme user or expert interviews; or group interviews. Observation techniques allow the team to learn from the users or stakeholders in their actual environment by observing (and maybe asking them questions), e.g. in their workplace. Observation techniques common in DT are Fly on the Wall, Shadowing or Contextual Inquiry. Immersion techniques aim for the team to make relevant experiences themselves if possible. These techniques include trying out a service or product by themselves, or simulating specific conditions, e.g. bad eyesight, for an extended period of time.

\subsection{Knowledge Sharing Techniques}
Knowledge sharing techniques support a DT team in distributing the information collected through research and testing activities.
A common technique for knowledge sharing is story telling.
In this activity, the involved researchers describe the details of their research to the rest of the team in the form of a story.
The other participants are asked to listens attentively and to collect the important information on sticky notes.
The form of a story should enable presenting the gathered information in a simple format and is supposed to provide reference points for anchoring key facts in the long-term memory of listeners.

\subsection{Knowledge Organization Techniques}
DT teams collect a vast amount of information that has to be organized in meaningful ways in order to learn from it. Common techniques to do so include clustering, Venn Diagrams, 2x2 Matrices, Timelines or process diagrams. With clustering, the team aims to group the information and looks for relationships or contradictions between clusters. Diagrams like matrices, Venn Diagrams or timelines help the team to organize their findings and highlight specific aspects, such as time or product categories and thereby uncover hidden needs and problems, or knowledge gaps.

\subsection{Knowledge Condensation Techniques}
In order to further work with the knowledge gained through research and testing, the team needs to condense it in such a way, that it can easily be accessed in later process stages. Personas, Point of View Statements, Storyboards, or Journey Maps are common techniques to condense knowledge. Personas represent a certain group of users with their needs, problems and wishes. The Point of View Statement condenses those problems from the problem domain the team currently wants to tackle. Storyboards or Journey Maps can be used to condense the experiences of many users into one typical experience.

\subsection{Idea Generation Techniques}
An integral part of every design process is to develop and discuss solution concepts. Many techniques from design or other creative disciplines can be used to develop ideas within a team. Common techniques include How to or How might we questions as brainstorming prompts, and a myriad of brainstorming techniques. The How to and How might we questions are often related to a Point of View Statement or a Persona derived from knowledge consolidation. They represent a specific problem to solve (How might we) or a specific aspect of the problem or the solution (How to). With regards to brainstorming techniques, different techniques work for different people and teams. Some brainstorming techniques let individuals brainstorm silently first and follow with a sharing and another group brainstorming session. Other brainstorming techniques use group brainstorming only. Some brainstorming techniques incorporate bodily activity. Some brainstorming techniques even use further prompts to generate ideas such as images or objects. Additionally DT prescribes a set of brainstorming rules.

\subsection{Prototyping Techniques}
In order to further develop an idea or test it with team members, users or stakeholders the idea needs to be prototyped. Prototypes can be a number of things, including simple sketches, product packages, roleplays, storyboards, wire frames, or functioning prototypes. Prototypes can have a different level of abstraction, ranging from low-fidelity prototypes designed to test basic interactions or ideas to high fidelity prototypes to test details of the interaction and functionality. Additionally, prototypes may cover the complete solution concept or single aspects of the concept.

\subsection{Testing Techniques}
Evaluating concepts with the actual user is an important aspect of DT. Several techniques to test concepts from areas such as UX Design, Human Computer Interaction, Design, and Software Development are part of the DT toolkit. Common techniques include the Think Aloud Method, Concept Testing, A/B Testing, or Usability testing. With the Think Aloud Method the user is asked to look at or try out the prototype and voice his thoughts out loud while doing so. The tester can ask clarifying questions. Concept Testing usually involves a low-fidelity prototype to explain the concept and an interview to gather feedback. A / B Testing is used to derive feedback for different alternatives be it features or concepts. Usability Testing aims to test how easy a product is to use.

\subsection{Feedback Techniques}
Feedback techniques can help to elicit and provide feedback, as well as document feedback. Most feedback techniques provide a specific set of categories in which to organize the feedback. For, example I like, I wish asks participants to provide feedback about what they liked and what they wish would have been different. Sometimes I like, I wish has an additional category called What if or How To in which participants can provide concrete examples to improve in the future. Other common feedback techniques include Five Finger Feedback, the Empathy Map, and the Feedback Capture Grid. 

\subsection{Facilitation Techniques}
Facilitation techniques aim to ensure conditions in which the team can effectively work towards their objectives. These techniques include those that help a team, manage conflicts, reflect on their process or manage their resources, e.g.time or energy. Common techniques include: Team Rules, which a team agrees on, Check-Ins that allow for team members to transition into the group work and let the team know what might affect their working today, Check-Outs or reflective sessions similar to retrospectives that provide space for teams to reflect on their process and their team work.
Voting techniques, such as dot voting or thermometer voting can help to arrive at a decision if no consensus can be reached.
Planning sessions help a team to decide on how to divide their time and work force. And the famous Time Timer helps keeping things time boxed.

\section{Mapping of Scrum Meetings and Techniques from the Design Thinking Toolkit }
By linking the collected requirements for each of the Scrum meetings and the goals and effects of the different types of techniques in the DT toolkit we derived a mapping between the two. Table \ref{table:mapping-theory} presents our theoretical mapping and the following sections provide explanations and examples for each entry.

\begin{table}[!t]
\caption{Mapping of Scrum meetings and Design Thinking techniques that could be applied to achieve the necessary meeting requirements and goals.}
\label{table:mapping-theory}

\renewcommand{\arraystretch}{1.3}
\centering
\begin{tabularx}{\columnwidth}{lX}
    \toprule
     \multicolumn{2}{p{0.95\columnwidth}}{\textbf{Theoretical mapping of Scrum meeting and applicable Design Thinking techniques}}\\
    \midrule
    General &
    \begin{itemize}[nosep,before=\leavevmode\vspace*{-1\baselineskip}] % [nosep]
        \item Warm-Ups 
        \item Facilitation techniques
    \end{itemize}\\
    \multicell{Product Backlog\\Refinement} &
    \begin{itemize}[nosep,before=\leavevmode\vspace*{-1\baselineskip}] % [nosep]
        \item Artifacts from knowledge organization techniques
        \item Artifacts from knowledge consolidation techniques 
        \item Artifacts from prototyping techniques
        \item Unpacking techniques 
        \item Idea generation techniques
        \item Low-fidelity prototyping techniques
        \item Voting techniques
    \end{itemize}\\
    Sprint Planning &
    \begin{itemize}[nosep,before=\leavevmode\vspace*{-1\baselineskip}] % [nosep]
        \item Artifacts from knowledge organization techniques
        \item Artifacts from knowledge consolidation techniques 
        \item Artifacts from prototyping techniques
        \item Unpacking techniques
        \item Idea generation techniques and their artifacts
        \item Voting techniques
    \end{itemize}\\
    Daily Scrum &
    \begin{itemize}[nosep,before=\leavevmode\vspace*{-1\baselineskip}] % [nosep]
        \item Unpacking techniques
        \item Knowledge sharing techniques
        \item Establishing and visualizing meeting rules
    \end{itemize}\\
    Sprint Review &
    \begin{itemize}[nosep,before=\leavevmode\vspace*{-1\baselineskip}] % [nosep]
        \item Testing techniques
        \item Feedback techniques
    \end{itemize}\\
    Retrospective &
    \begin{itemize}[nosep,before=\leavevmode\vspace*{-1\baselineskip}] % [nosep]
        \item Unpacking techniques
        \item Knowledge sharing techniques
        \item Feedback techniques
        \item Knowledge organization techniques
        \item Knowledge consolidation techniques
        \item Idea generation techniques
        \item Low fidelity prototyping techniques
        \item Retrospective as DT workshops 
    \end{itemize}\\
    \bottomrule
\end{tabularx}
\end{table}

\subsection{Techniques Generally Applicable to Scrum Meetings}
\label{subsec:generally_applicable}
DT warm-ups, short energizing activities, can be employed before longer meetings to motivate participants.
If several Scrum meetings are facilitated on the same day warm-ups can help the team transition from one meeting to the next and provide a welcome break. On such meeting days facilitation techniques such as check-ins and check-outs and establishing meeting rules can help the team transition into and out of the meeting work mode. Additionally the Time Timer can help staying within planned time boxes.

\subsection{Product Backlog Refinement}
During Refinement, developers need to understand the users' requirements, come up with solutions on how to implement these and document them. To support the understanding of the product vision as well as work items in general, artifacts from knowledge organization and consolidation techniques, which describe and document findings from user research, such as Personas, Journey Maps, Storyboards can be employed by the PO to help the team understand the underlying needs and business aspects. Similarly, artifacts from prototyping techniques such as wireframes and other low fidelity prototypes can be employed by the PO to help the team understand aspects of the functionality.
Additionally, when discussing how to implement a feature, unpacking methods such as charetting or creating a mind-map can help to form a shared understanding of the feature.
When discussing possible implementations, idea generation and prototyping techniques such as different types of brainstorming or sketching sessions can support the generation and discussion of possible solutions. The artifacts from these techniques, e.g. wire frames, diagrams, Journey Maps or Storyboards help to document the refined solution and can feed back to understanding the feature during planning or development.
Finally, voting techniques can support estimation and prioritization of work items.

\subsection{Sprint Planning}
The main goal of this meeting is to make a plan and commit to it.
As this is not a creative or problem-focused meeting, DT techniques are less helpful. However, some DT artifacts and techniques can support the meeting goals. As previously mentioned, several DT artifacts from knowledge organisation and condensation as well as from prototyping techniques can support the PO in explaining the necessary work and can help the team understand the needs, business aspects, and functionality behind the feature. Similarly, to the Product Backlog refinement, unpacking techniques help to achieve a shared understanding if the feature is not clearly enough defined yet. Additionally, voting techniques, such as Dot Voting or Thermometer Voting, can support the team in prioritizing and deciding which work items to include in the following sprint. And finally, the team can make use of idea generation techniques when formulating and discussing the sprint goal.

\subsection{Daily Scrum}
The daily Scrum, as the short status update is another meeting that is not creative and problem-focused. However, preparing the Daily Scrum meeting by quickly unpacking recent achievements and identified impediments as a group or as individuals can help stay in time. Establishing rules similar to DT's \emph{brainstorming rules} can ensure that discussions are deferred. A possible adaptation might be:
\begin{itemize}
    \item Defer judgment and discussions.
    \item Stay focused on the topic.
    \item One conversation at a time.
    \item Observers defer talking.
\end{itemize}
Additionally, these rules can be visibly placed around the meeting area as commonly seen with brainstorming rules in DT workspaces.

\subsection{Sprint Review}
As this meeting focuses on inspecting results and collecting feedback, actual customer representatives or stakeholders are present at these meetings who might not be familiar with the details relevant to software development.
DT testing techniques, such as the Think Aloud technique can support developers in eliciting targeted, actionable feedback.
Additionally, feedback techniques such as the Feedback Capture Grid, the Empathy Map or the Five Finger Feedback can support customers in giving structured feedback, which can be used to adapt work items or gauge the business value of features.

\subsection{Sprint Retrospective}
Creative activities for Retrospectives are already commonplace~\cite{jovanovic2016} and the toolbox of DT techniques can add to these. The Retrospective focuses on improving the work process and, as such, a large part of the DT process and its techniques can be applied to this meeting.

In fact, entire Retrospectives can be facilitated in the form of a DT workshop, including the understand or empathize step to collect attendees thoughts, a clustering step to identify major feedback items, a voting to decide which issues to work on for the next Sprint, an ideation phase to come up with solutions as well as a prototyping phase to sketch out the solution. A testing phase implements the solution in the following Sprint. 
During such a DT workshop as well as during "normal" retrospectives unpacking, knowledge sharing, and feedback techniques can support the team in reflecting on the past sprint. Knowledge organisation and condensation techniques help organize the feedback and identify improvements for the upcoming sprint. Idea generation and prototyping techniques can support the generation and documentation of action items.

\section{Concrete Examples from our Studies}

After building a theoretical mapping between the Scrum meetings and their requirements and different categories of techniques from the DT toolkit we analyzed the data from our two original studies in order to derive concrete examples from practice and compare them to our mapping. For each of the Scrum meetings Table \ref{table:mapping-practice} provides an overview of the DT techniques observed and mentioned in our study. As some of these techniques were not mentioned before, Table \ref{table:concrete_methods} briefly describes the specific DT techniques and artefacts mentioned in this section.

\begin{table}[!t]
\caption{Mapping of Scrum meetings and Design Thinking techniques based on the data from our study.}
\label{table:mapping-practice}

\renewcommand{\arraystretch}{1.3}
\centering
\begin{tabularx}{\columnwidth}{lX}
    \toprule
    \multicolumn{2}{p{0.95\columnwidth}}{\textbf{Practical experiences with Scrum meetings and applicable Design Thinking techniques}} \\
    \midrule
    General & 
        \begin{itemize}[nosep,before=\leavevmode\vspace*{-1\baselineskip}] % [nosep]
        \item Strict time boxing with a TimeTimer
    \end{itemize}\\
    Product Backlog Refinement &
    \begin{itemize}[nosep,before=\leavevmode\vspace*{-1\baselineskip}] % [nosep]
        \item Storyboards 
        \item Establishing and visualizing Personas
        \item Brainstorming to unpack features 
        \item Sketching / 30 seconds sketch  
    \end{itemize}\\
    Sprint Planning &
    \begin{itemize}[nosep,before=\leavevmode\vspace*{-1\baselineskip}] % [nosep]
        \item Storyboards 
        \item Establishing and visualizing Personas
        \item Otherwise not so useful 
    \end{itemize}\\
    Daily Scrum &
    \begin{itemize}[nosep,before=\leavevmode\vspace*{-1\baselineskip}] % [nosep]
        \item Nothing mentioned
    \end{itemize}\\
    Sprint Review &
    \begin{itemize}[nosep,before=\leavevmode\vspace*{-1\baselineskip}] % [nosep]
        \item Usability Testing
        \item Feedback Capture Grid
        \item Cooperative idea generation
    \end{itemize}\\
    Retrospective &
    \begin{itemize}[nosep,before=\leavevmode\vspace*{-1\baselineskip}] % [nosep]
        \item Retrospective as DT workshops 
        \item 3 Ls
        \item Clustering
        \item Brainstorming
        \item Storyboards
    \end{itemize}\\
    \bottomrule
\end{tabularx}
\end{table}

\begin{table}[!t]
\caption{Description of concrete DT methods and artifacts observed and mentioned in our studies.}
\label{table:concrete_methods}

\renewcommand{\arraystretch}{1.3}
\centering
\begin{tabularx}{\columnwidth}{lX}
    \toprule
    \textbf{Method} & \textbf{Description}\\
    \midrule
    Time Timer & \multicell{The Time Timer, a common tool to timebox DT workshops, is a clock\\ that visually displays remaining time. \textit{(facilitation)}}\\
    Storyboards & \multicell{Linear sequence of quickly drawn illustrations which describe a story.\\ \textit{(knowledge organisation, knowledge condensation, prototyping)}}\\
    Personas & \multicell{Fictional characters created for reference, based on user research, which\\ represent key user types. \textit{(knowledge condensation)}}\\
    30 Seconds Sketch & \multicell{ Effective technique to visualize and discuss a large quantity of ideas through\\ several 30 second rounds of sketching. \textit{(idea generation, prototyping)}}\\
    Usability Testing & \multicell{ Technique involving end-users in testing how easy a product can be used.\\ \textit{(testing)}}\\
    \multicell{Feedback Capture\\ Grid} & \multicell{Structured feedback collection using a 2x2 grid and the categories \emph{what}\\ \emph{worked}, \emph{what needs to change}, \emph{new ideas} and \emph{ open questions}. \textit{(feedback)}}\\
    3 Ls & \multicell{Structured feedback collection using the categories of \emph{liked}, \emph{learned} and\\ \emph{lacked}. \textit{(feedback)}}\\
    Brainstorming & \multicell{Spontaneously generating thoughts in relation to a specific promp, \\alone or in a group. \textit{(unpacking, idea generation)}} \\
    Clustering & \multicell{Organizing findings from research into named clusters and visualizing\\ relations between them. \textit{(knowledge organization)}}\\
    \bottomrule
\end{tabularx}
\end{table}

\subsection{Techniques generally applicable to Scrum Meetings}
The adherence to predefined time boxes for Scrum meetings can be supported by a common tool from DT workshops: the \emph{Time Timer}.
It is a large round analogue timer that visually displays the passage of time as well as the remaining time using a continuously decreasing red disk. After a developer experienced the Time Timer for the first time in a DT workshop he explained: \textit{''That was a real aha moment. If we could have these for our meetings that would be great.``}

\subsection{Product Backlog Refinement}\label{sec:PracticeBacklogRefinement}
One of the SMs from our study, who is also a developer, explained how he plans to introduce knowledge condensation techniques, to better describe features and achieve shared understanding when discussing backlog items: \textit{"We will start using Storyboards for larger features, explaining why the user needs this feature and how it is connected to the solution. Something we usually ignore, but with such Storyboards it can be done.
Storyboards can serve as the vision for this feature. Additionally, I would like to have Personas in the backlog for orientation and then sort the backlog around these Personas and stories for orientation instead of the 4 levels of items we have now.
Like this, we can achieve shared understanding early on."}
In a similar notion, another developer mentioned that he would like to have Personas around all the time to remember the users: \textit{"We should print these Personas and have them always near, so we never forget who is the user of the things we do."}
A developer explained how his team uses a quick form of brainstorming to discuss features and achieve a shared understanding: \textit{"If a task still has aspects that need clarification it can be interesting to do a short design thinking session. In the simplest case we just do a short brainstorming to discuss what the feature means for each of us."} 

Concerning idea generation techniques during refinement, developers saw potential in using ideation techniques, but not in all cases.
One developer stated: \textit{"With sketching, [he talked about the 30 seconds sketch technique] ]It is easier to communicate one's ideas, and seeing them drawn makes it easier to modify and build upon them}. However, another developer noted: \textit{It depends on the level of refinement. If all that remains are technical challenges, DT does not really help. If aspects of the feature are not clear yet then DT is a really interesting approach to refinement."}

\subsection{Sprint Planning}
In our study, the process of planning the next items to work on was rarely mentioned in conjunction with DT techniques. As described in \ref{sec:PracticeBacklogRefinement}, one interviewee wanted to use storyboards and personas whenever discussing backlog items. In contrast, other study participants pointed out that the Scrum planning meeting relied on facts, such as the velocity of the team, the amount of work they forecast could be completed in the upcoming development iteration and the priority of work items. As one developer explained: \textit{''For planning it's not a fit I think. It's about prioritizing and the capacity of the team; it's just not a creative process.``}

\subsection{Daily Scrum}
The Daily Scrum meeting, as a regular check-in and team synchronization meeting, was not explicitly mentioned by interviewees as requiring the support of DT techniques.
However, Daily Scrum meetings should ideally be short and rely on strict time-boxes. As described in Section~\ref{subsec:generally_applicable} the Time Timer can help teams with time keeping especially during this meeting. 

\subsection{Sprint Review}
The use of testing techniques during review meetings to collect feedback from customers and stakeholders were mentioned several times in our studies.
For example, one interviewee noted: \textit{''Evaluation techniques --- not traditional ones, where you ask questions and they answer, but for example, you just give the system to the user and see how he interacts with it --- that might help in getting better feedback.``} 

Furthermore, feedback techniques were mentioned as being useful.
One developer explained how a feedback capture grid not only helped to collect feedback but also led to cooperative ideation with the customer: \textit{''I think it was good, and it helps to share different point of views. You can see what is liked and disliked, and the criticisms makes you ask for questions and the questions gives you some ideas. So there is a cycle for generating ideas.``}

\subsection{Sprint Retrospective}
One of the SMs from our studies facilitates his Retrospectives as DT workshops: \textit{"Running a feedback session with 30 people is chaos, it would be a whole day retro with nothing achieved. For large groups more organization is needed so we collected feedback with the 3 Ls in the group, quickly clustered and then split up into smaller groups for the next steps. That's where DT really makes sense."} As next steps, the smaller groups in this retrospective would brainstorm solutions and prototype them in a story board or sketch.
A developer from his team describes his first experience with such a DT-Retrospective as follows: \textit{"That big retro was a very nice experience. It was totally different from what we are used to. Normally, we just sit in a room, but here I find it was very interactive and I liked that people were open. I felt that most of the conclusions from that meeting are on their way now, so it really had a good impact on our team."}

\section{Discussion}
In this chapter, we analyzed the integration of DT techniques into the meeting activities of the Scrum software development method from two perspectives.

First, we analyzed the five central Scrum meetings and reviewed their goals, requirements, and outcomes along with categories of DT techniques and the tasks they support.
From the outcomes of this step, we derived a mapping of Scrum meetings and applicable DT techniques.
In this theoretical mapping, we showed links between at least two categories of DT techniques with each of the five investigated Scrum meetings.
We found the most matches between categories of DT techniques and the Retrospective, the Product Backlog Refinement, as well as the Sprint Planning meetings.
This finding is in line with related research and practice for Scrum activities in meetings, which also concentrate on Retrospectives, Product Backlog Refinement and Sprint Planning.

In a second step, we analyzed data from two previous studies in which the topic of \textit{DT techniques supporting Scrum meetings} emerged.
This meta-analysis provided concrete examples from practice of how DT techniques support the facilitation of Scrum meetings.
In our case study, we found examples of DT techniques supporting four out of the five Scrum meetings.
For the remaining meeting, the Daily Scrum we could not observe any specific methods and the participants did not specifically mention the meeting in our interviews. For the Sprint Planning, some study participants specifically mentioned that they do not consider DT techniques useful for this event, while others considered artifacts from knowledge condensation within the backlog as useful to establish vision and shared understanding.

\begin{table}[!t]
\caption{Mapping of Scrum meetings and Design Thinking techniques highlighting our evaluation results (light gray - mappings that could be confirmed with our data}
\label{table:mapping-comparison}

\renewcommand{\arraystretch}{1.3}
\centering

\begin{tabularx}{\columnwidth}{lX}
    \toprule
     \multicolumn{2}{p{0.95\columnwidth}}{\textbf{Comparison of theory and practice of Scrum meetings and applicable Design Thinking techniques}}\\
    \midrule
    General &
    \begin{itemize}[nosep,before=\leavevmode\vspace*{-1\baselineskip}] % [nosep]
        \item Warm-Ups
        \item \color{black}\colorbox{lightgray}{Facilitation techniques (timeboxing with TimeTimer)}
    \end{itemize}\\
    \multicell{Product\\Backlog\\Refinement} &
    \begin{itemize}[nosep,before=\leavevmode\vspace*{-1\baselineskip}] % [nosep]
        \item Artifacts from knowledge organization techniques
        \item \colorbox{lightgray}{Artifacts from knowledge consolidation techniques (persona, storyboard)} 
        \item Artifacts from prototyping techniques 
        \item \colorbox{lightgray}{Unpacking techniques (brainstorming)}
        \item \colorbox{lightgray}{Idea generation techniques (30 seconds sketch)}
        \item \colorbox{lightgray}{Low-fidelity prototyping techniques (sketching)}
        \item Voting techniques
    \end{itemize}\\
    Sprint Planning &
    \begin{itemize}[nosep,before=\leavevmode\vspace*{-1\baselineskip}] % [nosep]
        \item Artifacts from knowledge organization techniques
        \item \colorbox{lightgray}{Artifacts from knowledge consolidation techniques (persona, storyboard)}
        \item Artifacts from prototyping techniques
        \item Unpacking techniques
        \item Idea generation techniques and their artifacts
        \item Voting techniques
    \end{itemize}\\
    Daily Scrum &
    \begin{itemize}[nosep,before=\leavevmode\vspace*{-1\baselineskip}] % [nosep]
        \item Unpacking techniques
        \item Knowledge sharing techniques
        \item Establishing and visualizing meeting rules
    \end{itemize}\\
    Sprint Review &
    \begin{itemize}[nosep,before=\leavevmode\vspace*{-1\baselineskip}] % [nosep]
        \item \colorbox{lightgray}{Testing techniques (usability testing)}
        \item \colorbox{lightgray}{Feedback techniques (feedback capture grid)}
    \end{itemize}\\
    Retrospective &
    \begin{itemize}[nosep,before=\leavevmode\vspace*{-1\baselineskip}] % [nosep]
        \item Unpacking techniques
        \item Knowledge sharing techniques
        \item \colorbox{lightgray}{Feedback techniques (3 Ls)}
        \item \colorbox{lightgray}{Knowledge organization techniques (clustering)}
        \item Knowledge consolidation techniques
        \item \colorbox{lightgray}{Idea generation techniques (brainstorming)}
        \item \colorbox{lightgray}{Low fidelity prototyping techniques (storyboards)}
        \item \colorbox{lightgray}{Retrospective as DT workshops}
    \end{itemize}\\
    \bottomrule
\end{tabularx}
\end{table}

Comparing the two created mappings, we find several of our theoretical links confirmed by concrete examples from practice. Table \ref{table:mapping-comparison} provides an overview of which entries could be confirmed and which entries could not be confirmed within this study.
However, some of the categories of DT techniques were not mentioned or observed in our studies. For example, the anticipated general applicability of DT warm-up techniques for all Scrum meetings could not be confirmed as part of this study. This might be explainable by the perceptions of participants. Some of the better known warm ups and activities, especially those that include elements of playfulness, are easily perceived as childish games or ``fluff'' by professionals~\cite{Verity2012,Kolko2018}.

A noticeable difference between the mappings can be seen for the Sprint Planning meeting. While the meeting requirements, goals, and outcomes match with several DT activities the data from our studies only supports one of these matches: the use of knowledge condensation techniques to explain the vision of a feature and support shared understanding. Additionally participants even mentioned that they don't think DT activities are useful in this meeting. A possible explanation for this discrepancy might be found in the different ways planning meetings are facilitated. While in some teams Sprint Planning and Backlog Refinement are held as one meeting or as two meetings directly following each other, other teams separate the two meetings by a number of days. Thus, a Sprint Planning meeting which directly follows a Backlog Refinement probably focuses on planning and prioritizing action items and does not require further establishment of shared understanding. For a Sprint Planning that does not include or follow a Backlog Refinement the theoretically useful artifacts and techniques might be more relevant.

Another interesting finding was the mention of the 3Ls as a DT technique. This technique is commonly known as a retrospective activity for the data gathering stage, albeit as the 4Ls \footnote{4Ls Retro activity https://retromat.org/de/?id=78}. From our experience this activity is less common in the DT community. While it is hard to say in which area activities or techniques were first introduced and used, we believe this is an indicator that the toolkit of Scrum meetings and the DT toolkit are compatible and seem to have merged in the eyes of the interviewee.

\section{Conclusion}

This study is based on two previous studies investigating DT use in software development teams but did not specifically focus on DT techniques for Scrum meetings.
Accordingly, these previous studies featured an adequate sample of participating teams but provided only a small selection of concrete examples for DT techniques used in Scrum meetings.
Further studies explicitly focused on this topic, can confirm additional entries in our mapping or can add additional new entries and concrete examples.
Additionally, in this study, we investigated the five central Scrum meetings, which are generally applied in professional software development contexts.
Several additional meetings exist, especially in different scaled versions of Scrum.
One such example is the \emph{Scrum of Scrums} meeting, introduced by one of the originators of the Scrum method~\cite{Sutherland2001}, which provides a meeting with members of multiple development teams.
Future work should, therefore, investigate the use of DT techniques in these scaled Scrum meetings.

Despite these limitations, our results indicate that integrating DT methods into Scrum meetings is a promising area of research with practical implications.
Several of the DT activities that can be added to the toolbox of Scrum activities promote collaboration and a shared understanding within the development team. They help achieve meeting goals and prevent common Scrum meeting issues. As added activities to the Scrum toolkit they provide a wider variety of methods for each Scrum meeting, promote diversity, and prevent monotonous meetings.
Finally, development teams are confronted with additional use cases for DT techniques they might already be familiar with from DT workshops or DT project phases. Thus Scrum team members have the opportunity to further practice these DT techniques.
However, not all Scrum meetings can or should be supported by DT in the same way, i.e., Sprint Planning and Daily Scrum can only marginally integrate DT techniques, while the Retrospective and the Product Backlog Refinement provide several opportunities to make use of DT techniques.

This research complements existing efforts to integrate DT and agile software development methods.
Our results provide a starting point for practitioners to leverage DT's advantages, not only in an initial DT phase but also during regular development activities, in a similar fashion as proposed in the Ad-hoc DT model or the Toolbox model.
The mapping developed in this chapter presents an initial guideline on how to integrate individual DT techniques into existing Scrum meetings.
It provides motivation and ideas on enhancing Scrum meeting routines with further techniques for meeting facilitators and teams: either identifying additional use cases for already familiar techniques or by pointing out new techniques worth investigating.

%%%%%%%%%%%%%%%%%%%%%%%% referenc.tex %%%%%%%%%%%%%%%%%%%%%%%%%%%%%%
% sample references
% %
% Use this file as a template for your own input.
%
%%%%%%%%%%%%%%%%%%%%%%%% Springer-Verlag %%%%%%%%%%%%%%%%%%%%%%%%%%
%
% BibTeX users please use
\bibliographystyle{unsrt}
\bibliography{bib}
\end{document}